\documentstyle[prl,multicol,aps,psfig]{revtex}
\def \beq{\begin{equation}}
\def \eeq{\end{equation}}
\def \beqa{\begin{eqnarray}}
\def \eeqa{\end{eqnarray}}
\def \ppbar{\langle\overline\psi\psi\rangle}
\begin{document}
\draft
\title{Quenched QCD at finite temperature with chiral Fermions}
\author{Rajiv V.\ Gavai \cite{ervg} and Sourendu Gupta \cite{esg}}
\address{Department of Theoretical Physics, Tata Institute of Fundamental
         Research,\\ Homi Bhabha Road, Mumbai 400005, India.}
\author{R.\ Lacaze \cite{erl}}
\address{Service de Physique Theorique, CEA Saclay,\\
         F-91191 Gif-sur-Yvette Cedex, France.}
\maketitle
\begin{abstract}
We study physics at temperatures just above the QCD phase transition
($T_c$) using chiral (overlap) Fermions in the quenched
approximation of lattice QCD. Exact zero modes of the overlap Dirac
operator are localized and their frequency of occurrence drops with
temperature. This is closely related to axial $U(1)$ symmetry, which
remains broken up to $2T_c$. After subtracting the effects of these
zero modes, chiral symmetry is restored, as indicated by the behavior
of the chiral condensate ($\ppbar$). The pseudoscalar and vector
screening masses are close to ideal gas values.

\end{abstract}
\pacs{11.15.Ha, 12.38.Mh\hfill TIFR/TH/01-27, t01/076, hep-lat/0107022}

\begin{multicols}{2}

With new results from the Brookhaven heavy-ion collider appearing thick
and fast \cite{qm01}, the time seems ripe for making a concerted effort
to understand the dynamics of the high-temperature phase of quantum
chromodynamics, namely the quark-gluon plasma.  There are several puzzles
that seem to have resisted a decade of efforts to understand them. The
one we focus on involves the static screening of certain excitations of
the plasma.

It has long been understood that the screening of currents in a plasma
would give us information on its excitations. Currents with certain
quantum numbers excite mesons from the vacuum at low temperatures,
and should exhibit deconfinement related changes above the QCD phase
transition temperature ($T_c$) \cite{detar}. Detailed studies have shown
that this indeed does happen in the vector, and axial-vector channels: the
screening above $T_c$ is clearly due to nearly non-interacting quark
anti-quark pairs in the medium \cite{mtc,tifr}. On the other hand,
the scalar and pseudo-scalar screening masses show more complicated
behavior--- strong deviations from the ideal Fermi gas, and a strong
temperature dependence.  This puzzling behavior is generic--- it has
been seen in quenched \cite{quenched} and dynamical simulations with
two \cite{nf2} and four flavors \cite{detar,mtc,tifr,nf4} of staggered
quarks, as well as with Wilson quarks \cite{wilson}. This is the puzzle
that we address and solve in this letter.

The new technique we bring to bear on this problem is to use a version
of lattice Fermions called overlap Fermions \cite{neu}. It has the
advantage of preserving chiral symmetry on the lattice for any number of
massless flavors of quarks \cite{luescher}.  This is in contrast to other
formulations such as Wilson Fermions which break all chiral symmetries
or staggered Fermions which break them partially.  Since the number of
pions and their nature is intimately related to the actually realized
chiral symmetry on the lattice, we should expect any realization of chiral
Fermions on the lattice to provide insight into the question we address.

The overlap Dirac operator ($D$) can be defined \cite{neu2} in terms of the
Wilson-Dirac operator ($D_w$) by the relation
\beq
   D = 1 - D_w (D_w^\dag D_w)^{-1/2}.
\label{overlap}\eeq
The computation of $D^{-1}$ needs a nested series of two matrix inversions
for its evaluation (each step in the numerical inversion of $D$ involves
the inversion of $D_w^\dag D_w$).  This squaring of effort makes a
study of QCD with dynamical overlap quarks very expensive. As a first
step in this direction, we chose to work with quenched overlap quarks:
to study the pattern of chiral symmetry restoration and screening masses
at high temperature.

We generated quenched QCD configurations at temperatures of $T/T_c$ =
1.25, 1.5 and 2 on $4 \times 8^3$ and $4 \times 12^3$ lattices (see Table
\ref{tb.summary}).  The corresponding couplings are respectively $\beta=
$ 5.8, 5.8941 and 6.0625.  The configurations were separated by 1000
sweeps of a Cabibo-Marinari update.  A previous computation \cite{liu}
with $T=0$ quenched overlap quarks at the nearby couplings of $\beta=5.85$
and 6.0 allows us to compare finite and zero temperature physics.

For the matrix $M=D_w^\dag D_w$, and a given source vector $b$,
we computed $y=M^{-1/2} b$ by a conjugate gradient (CG) version
of a proposed Lancz\"os method \cite{borici}. CG gives better
control over errors than Lancz\"os or other methods based on
approximations using Chebychev or Reeves polynomials.  In CG,
the vector $x=M^{-1}b$ is obtained by iterating the residual
vector $r_{i+1}=r_i-\alpha_i M p_i$ and the direction of change
$p_{i+1}=\beta_{i+1} p_i + r_{i+1}$, where $\alpha_i=(r_i,r_i)/(p_i,M
p_i)$ and $\beta_{i+1}=(r_{i+1},r_{i+1})/(r_i,r_i)$.  The iterations
are stopped when $(r_i,r_i) <\epsilon$, for a predetermined tolerance
$\epsilon$.  Let $N_s$ be the number of CG-iterations at stopping.
In the orthonormal basis made of $q_i=r_i/(r_i,r_i)^{1/2}$, the
matrix $M$ can be approximated by $Q^\dagger {\cal T} Q$, where
$Q$ is made from the set \{$q_i$\} and ${\cal T}$ is a tridiagonal
symmetric matrix of dimension $N_s$. The non-zero elements of ${\cal T}$ are
${\cal T}_{i,i} = 1/\alpha_{i} +\beta_i/\alpha_{i-1}$ and ${\cal
T}_{i,i+1}=-\sqrt{\beta_i}/\alpha_{i-1}$.  Denoting by $\Lambda$
the diagonal matrix of the eigenvalues of ${\cal T}$, and by $U$ its
diagonalizer made from the corresponding eigenvectors, the desired
solution is $y = M^{-1/2}b \simeq Q^\dagger U^\dagger \Lambda^{-1/2} U Q b$,
when the CG iterations are started with $r_0=p_0=b$.

\begin{figure}[htb]\begin{center}
   \leavevmode
   \psfig{figure=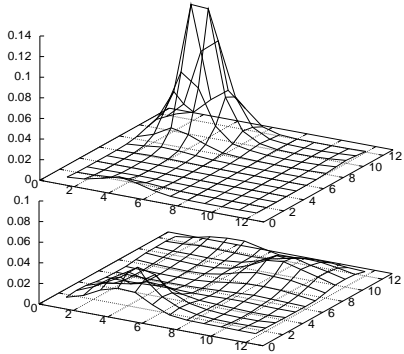,height=5.2cm,width=6cm}
   \end{center}
   \caption{Projection on the $xy$ plane of two typical eigenvectors
    on a $4\times12^3$ lattice at $T=1.5 \ T_c$. The zero eigenvector
    (top) is strongly localized while the non-zero eigenvector (bottom)
    is not.}
\label{fg.local}\end{figure}

A massive overlap operator is defined by
\beq
   D(ma) = ma + (1-ma/2) D,
\label{moverlap}\eeq
where $m$ is the bare quark mass, $a$ the lattice spacing, and $D$
is defined in (\ref{overlap}).  We used the usual quark propagator,
$G(ma)=[1-D/2] D^{-1}(ma)$ \cite{neu2}.  We computed $G$ on 12 point
sources (3 colors and 4 spins) for 10 quark masses from $ma$=0.001 to
0.5 using a multimass inversion of $D^{\dagger}D$. The (negative) Wilson
mass term in $D_w$, which is an irrelevant regulator, was set to 1.8.
The tolerance was $\epsilon=10^{-6}$ in the inner CG and $10^{-4}$ in
the outer CG.  A rough computation of the eigenvalues of $D^{\dagger}D$,
$\mu^2$, was made on each configuration with a Lancz\"os method in
each chiral sector.  Whenever $\mu^2 \simeq O(10^{-5})$ was obtained,
the few lowest eigenvalues and eigenvectors were refined to a precision
of about $10^{-8}$ by a Ritz functional minimization.

For each configuration we verified that the Ginsparg-Wilson \cite{gw}
relation is satisfied to an accuracy of $10^{-9}$, thus ensuring a
very precise implementation of chiral symmetry for our simulations.
Another test of the precision of our measurements was provided by a check
of the chiral Ward identity---
\beq
   a^2\chi_{PS} = {1\over{ma}}a^3\ppbar,
\label{gor}\eeq
where $\ppbar$ is the chiral condensate and $\chi_{PS}$ is the
pseudoscalar susceptibility \cite{note1}.  In all our computations we
found that the equality was satisfied to better than 1 part in $10^5$.

For most configurations we found that the spectrum of $D^\dag D$
starts well away from zero. However, for some configurations we
found zero and near-zero modes with $\mu^2\lesssim10^{-4}$ clearly
seperated by a gap from the non-zero modes with $0.1\lesssim\mu^2$
\cite{note0}. The non-zero modes clearly came in degenerate pairs of
opposite chiralities. For $T=1.5T_c$ and $2T_c$ the zero modes were all
less than $10^{-7}$ and of definite chirality. Their numbers decreased
rapidly with either increasing $T$ or decreasing lattice size (see Table
\ref{tb.summary} for details).  We constructed a gauge-invariant measure
of localization \cite{gockeler} for a normalised eigenvector $\Phi(i)$
where $i$ stands for position, spin and color, as the following sum,
\beq
   \sigma = \sum_{\rm sites} \left[\sum_{\rm spin,color}
                {\overline\Phi}(i)\Phi(i)\right]^2.
\label{local}\eeq
This varies from unity for an eigenmode localized at just one site to
$1/V$ for an eigenmode spread uniformly on a lattice of volume $V$. On
$4\times12^3$ lattices we found $\sigma\simeq(3-8)\times10^{-3}$
for the zero modes and $\sigma<10^{-3}$ for the non-zero modes (see
Figure \ref{fg.local}).  At $T=1.25T_c$ we also found near-zero modes
with $\mu^2\lesssim10^{-4}$.  These were as localised as the zero modes,
but came paired in parity like the non-zero modes.

The zero modes of chiral Fermions are related to instanton-like
configurations \cite{instanton}. These are, in turn, known to break
axial $U(1)$ symmetry.  For two flavors, the order parameter for the
axial $U(1)$ symmetry \cite{ids} is the difference of the flavor singlet
and triplet scalar susceptibilities---
\beq
   \omega = \chi^3_S - \chi^0_S = \frac{4}{m^2}
            \frac{\langle(n_+-n_-)^2\rangle}V\,,
\label{u1a}\eeq
where $n_+$ and $n_-$ are the number of eigenvalues of $D$ with left
and right handed chiralities, and the difference needs to be evaluated
only for the zero modes. In the quenched theory $\omega$, if non-zero,
is singular in the $m\to0$ limit since $\langle(n_+-n_-)^2\rangle$
is independent of $m$.  This is related to well known problems with
$\chi^0_S$ in the quenched theory \cite{qclogs}.  Combining our data with
\cite{heller}, we find some evidence that $\langle(n_+-n_-)^2\rangle/V$
falls as a high power of $T/T_c$. Nevertheless in quenched QCD, $U(1)$
symmetry is not completely restored even at $2T_c$.

Our measurement of $\ppbar$ comes from the diagonal part of
$G(ma)$ on the 12 sources used for each configuration. The analysis
proceeds by writing $G$ in terms of the eigenvectors $\Phi^\mu_\alpha$
of $D^\dag D$ with eigenvalue $\mu^2$ and chirality $\alpha$,
\beq
   G_{ij}=\sum_{\mu}\Phi^{\mu}_{\alpha}(i) {\overline \Phi}^{\mu}_{\beta}(j)\;
       {ma\lambda^2\delta_{\alpha,\beta}-i\mu\lambda(1-\delta_{\alpha,\beta})
          \over{\mu^2+(ma)^2\lambda^2}}\;,
\label{quark}\eeq
where $\lambda^2=1-\mu^2/4$.  The contribution of a zero mode can be
easily read off from the equation above and is seen to be proportional
to $\Phi\overline\Phi$ and $1/ma$. Since the eigenvector corresponding
to the zero mode is localized, its contribution to the condensate depends
strongly on the spatial position of the source vector. The remaining modes
are delocalized and closely spaced; so the overlap of the eigenvector
on the source is averaged out.

\begin{figure}[htb]\begin{center}
   \leavevmode
   \psfig{figure=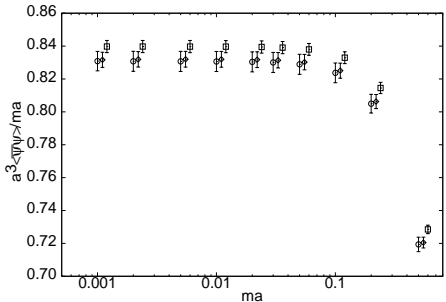,height=4.2cm,width=6cm,angle=-90}
   \end{center}
   \caption{$a^3\ppbar/ma$ as a function of $ma$ at $T=1.25T_c$ (circle),
      $1.5T_c$ (diamond) and $2T_c$ (square) on $4\times12^3$ lattices. 
      Zero and near-zero mode contributions are subtracted. Data for
      $1.5T_c$ and $2T_c$ are displaced in $ma$ for visibility.}
\label{fg.cond}\end{figure}

Our precise determination of the eigenvectors and eigenvalues allows us
to subtract out the zero mode contributions in (\ref{quark}), although
it can sometimes be a couple of orders of magnitude larger than the
remainder. The subtracted condensate is strikingly identical to that
seen in the sample without zero modes, at all the couplings and lattice
sizes studied.  We found that $\ppbar$ varies linearly with $ma$ and goes
to zero as $ma\to0$.  This is how chiral symmetry restoration manifests
itself in quenched QCD.

In the thermodynamic and continuum limits, it is not clear whether the
near-zero modes are related to chiral symmetry breaking; on any finite
$V$ they are not, but in these limits they may accumulate at zero.
If they do, then $\ppbar$ would not go to zero with $m$ even above
$T_c$ in the quenched theory. Clearly, one needs to examine these
limits very carefully.  At $T=1.25T_c$ the number of near-zero modes is
insignificant, but it seems that for $T < 1.25T_c$ their numbers will be
larger \cite{heller} and the $m\to0$ limit will have to be taken after
the $V\to\infty$ limit. Quite likely, such a limit may have to be taken
at more than one lattice spacing.

For lattice Fermions which satisfy the Ginsparg-Wilson
relation \cite{gw}, the following identities hold
in the chirally symmetric phase as $ma\to0$,
\beq
   C_S(z) = -C_{PS}(z) \quad{\rm\ and}\quad C_V(z)=C_{AV}(z).
\label{corid}\eeq
Here $C$ is the screening correlation function in the spatial
$z$ direction of an operator summed in the other three directions.
The subscripts $PS$ refer to a pseudo-scalar operator,
$S$ to a scalar, $V$ to a vector and $AV$ to an axial-vector \cite{note2}.
We find that the V and AV correlators indeed agree at all
$z$ and at all temperatures we studied.  This is in agreement with our
conclusion that the temperature range we studied has chiral symmetry.

\begin{figure}[htb]\begin{center}
   \leavevmode
    \psfig{figure=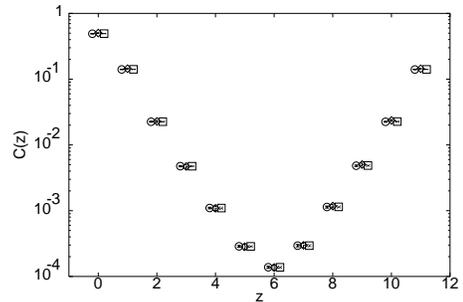,height=4.2cm,width=6cm,angle=-90}
   \end{center}
   \caption{The correlators $C_{PS}$ for configurations without zero modes
      (circle), ${\overline C}_{PS}$ for configurations with zero modes
      (diamond) and $(C_{PS}-C_S)/2$ for all configurations (square)
      for $4\times12^3$ lattices with $ma=0.001$ and $T=1.5T_c$. Some of the
      data points are displaced in $z$ for visibility.}
\label{fg.zeromodes}\end{figure}

It is clear from the chiral Ward identity (\ref{gor}) that large
fluctuations in $\ppbar$ due to the zero mode must also lead to similiar
non-statistical fluctuations in $C_{PS}$.  Further, since a zero
mode contributes identically to $C_S$ and $C_{PS}$, the two can even
have the same sign if this contribution is large enough \cite{note3}.
Ignoring the configurations with zero modes, the S and PS
correlators do obey the identity in (\ref{corid}), suggesting that
simple results could be seen by eliminating the zero mode contribution.

One way to do this is to use the measured $\Phi$'s to subtract
the zero modes and construct a new correlator ${\overline
C}_{PS}(z)$. Alternatively, we could consider the difference,
$(C_S(z)-C_{PS}(z))/2$, which should equal ${\overline C}_{PS}(z)$.
Figure \ref{fg.zeromodes} exhibits the comparison of the three
correlators.  Similar excellent agreement is also seen for these
correlators for all $ma<0.1$ and at other $T$.

\begin{figure}[htb]\begin{center}
   \leavevmode
    \psfig{figure=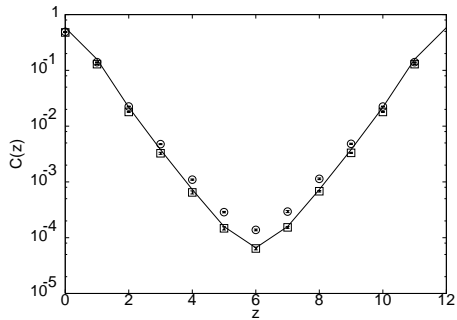,height=4.2cm,width=6cm}
   \end{center}
   \caption{The screening correlators on $4.12^3$ lattices at
      $T=1.5T_c$ for $ma=0.001$, in analyses without zero modes. The
      V/AV correlators (boxes) agree very well with the ideal gas
      computation (solid line), and the S/PS correlators (circles)
      are also similar.}
\label{fg.ideal}\end{figure}

After subtracting the effects of the zero mode, we find that the
correlator identities are satisfied for the S/PS sector as well as
V/AV. In addition, $C_V$ is described well by an ideal gas of overlap
quarks on the same lattice, as illustrated in Figure \ref{fg.ideal}
(although the figure shows this only for one quark mass and $T$, this
is true for all $T\ge1.25 \ T_c$ and all $ma<0.1$). While $C_{PS}$
seems to differ from the ideal gas result, the measured values of the PS
screening masses, $M_{PS}$, are only 10\% smaller than ideal gas screening
masses. This difference is small enough to be plausibly explained in a
weak-coupling computation, quite unlike earlier results from staggered
or Wilson quarks.

Differences between $T=0$ mesons and our measurements are extremely
clear.  For $T\ge1.25T_c$, $M_{PS}$ is constant and non-vanishing
for $ma\lesssim 0.1$. In the same temperature range, the ratio
$m_{PS}(T)/m_V(T)$ is within 10\% of unity and quite different from the
measured values at $T=0$ at nearby couplings $\beta$ \cite{liu}. Thus,
the simple picture that emerges is a property of the high temperature
phase of (quenched) QCD.

In conclusion, working with chiral (overlap) Fermions, we have found
several new results and a consistent picture of the high temperature
phase of quenched QCD. Axial $U(1)$ symmetry is not restored even at
$2T_c$. As a result the thermal ensemble contains gauge fields which
give rise to Fermion zero modes of definite chirality. When the effect
of these modes is subtracted, $\ppbar$ vanishes in the zero quark
mass limit, showing that chiral symmetry is restored. Simultaneously,
parity doubling is seen in the spectrum of screening masses, which are
close to those expected in an ideal Fermi gas, even for the S/PS sector.
Since some of these results are not obtained with staggered quarks, it
is an interesting question whether the two flavor QCD phase transition
is properly described by such a representation of quarks.

Some interesting problems remain to be solved.  At $T\le1.25\ T_c$, there
are near-zero modes.  It cannot be ruled out that these modes shift the
quenched chiral symmetry restoration point away from $T_c$.  However,
this question is crucially related to the evolution of near-zero modes
with lattice volume and spacing. Hence the nature of these complications,
and the question of whether they are quenched artifacts or remain in
full QCD, will only become clear with further studies which are underway.

This work was funded by the IFCPAR as its project 2104-2.

\begin{table}[hbtp]\begin{center}
  \begin{tabular}{crrccccc}  \hline
  $T/T_c$ & $V$ & $N_c$ & $N_0$ & $N_0'$ & $\ppbar/(mT^2)$ &
            $M_{PS}/T$ & $M_V/T$ \\
  \hline
  $1.25$ & $4\times12^3$ &  50 & 18 & 4 & 13.24 (9) & 6.0 (2) & 6.6 (1) \\
  $1.50$ & $4\times12^3$ &  50 &  8 & 0 & 13.31 (8) & 6.0 (1) & 6.5 (1) \\
  $2.00$ & $4\times12^3$ &  50 &  1 & 0 & 13.44 (6) & 6.0 (1) & 6.5 (1) \\
  \hline
  $1.50$ &  $4\times8^3$ & 100 &  1 & 0 & 13.36 (5) & 6.0 (2) & 6.5 (2) \\
  $2.00$ &  $4\times8^3$ &  26 &  0 & 0 & 13.34 (8) & 6.0 (1) & 6.5 (1) \\
  \hline
  \end{tabular}
  \end{center}
  \caption[dummy]{Temperatures ($T$), the lattice volume ($V$), and the number
     of configurations analysed ($N_c$). Also shown are the number of
     zero modes ($N_0$) and near-zero modes ($N_0'$), and $\ppbar/m$, and the
     PS and V screening masses for $m\to0$.}
\label{tb.summary}\end{table}

\end{multicols}

\begin{thebibliography}{99}
\bibitem{note1}
  ``Meson'' susceptibilties are defined to be the zero 4-momentum
  component of the corresponding correlator.
\bibitem{note0}
  The edge of the band of zero and near-zero modes is at $|\mu|\approx10$ MeV,
  for $T=1.25T_c$ and $T_c\approx300$. There was also a seperation of two
  orders of magnitude between the zero and near-zero modes.
\bibitem{note2}
  In order to remove intermediate vacuum insertions we
  take $C_S(z)$ to be an iso-vector correlator.
\bibitem{note3}
  Zero modes do not contribute to the $V$ and $AV$ correlators unless they
  appear with both chiralities on a given configuration. Even in that case
  the contribution is significant only if they are also located near each
  other.
\end{thebibliography}
\end{document}